\documentclass[aps,prl,reprint,superscriptaddress]{revtex4-2}
\usepackage{amsmath}
\usepackage{amssymb}
\usepackage[utf8]{inputenc}
\usepackage{siunitx} 
\usepackage{xcolor}
\usepackage{natbib}
\usepackage{graphicx}
\usepackage{braket}
\usepackage{comment}
\usepackage{bm}
\usepackage{epstopdf}
\begin{document}

%\title{ZnO QDs}
% BELLOW IS THE OLD TITLE
%\title{Correlated electronic structure theory of II-VI QDs}

% BELLOW IS THE NEW TITLE THAT EXPLAINS CONTENT OF THE MANUSCRIPT PERHAPS BETTER
%\title{Interplay between multipole expansion of exchange interaction and Coulomb correlation for exciton fine structure of colloidal II-VI QDs}

\title{Interplay between multipole expansion of exchange interaction and Coulomb correlation of exciton in colloidal II-VI quantum dots}

\date{\today}

\author{Petr Klenovsk\'y}
\email{klenovsky@physics.muni.cz}
     \affiliation{Department of Condensed Matter Physics, Faculty of Science, Masaryk University, Kotl\'a\v{r}sk\'a~267/2, 61137~Brno, Czech~Republic}
     \affiliation{Czech Metrology Institute, Okru\v{z}n\'i 31, 63800~Brno, Czech~Republic}

\author{Jakub Valdhans}
     \affiliation{Department of Condensed Matter Physics, Faculty of Science, Masaryk University, Kotl\'a\v{r}sk\'a~267/2, 61137~Brno, Czech~Republic}

\author{Lucie Krej\v{c}\'{i}}
     \affiliation{Department of Condensed Matter Physics, Faculty of Science, Masaryk University, Kotl\'a\v{r}sk\'a~267/2, 61137~Brno, Czech~Republic}

\author{Miroslav Valtr}
%\email{klapetek@cmi.cz}
     \affiliation{Czech Metrology Institute, Okru\v{z}n\'i 31, 63800~Brno, Czech~Republic}
     \affiliation{CEITEC, Brno University of Technology, Purkyňova 123, 612 00 Brno, Czech Republic}

\author{Petr Klapetek}
%\email{klapetek@cmi.cz}
     \affiliation{Czech Metrology Institute, Okru\v{z}n\'i 31, 63800~Brno, Czech~Republic}
     \affiliation{CEITEC, Brno University of Technology, Purkyňova 123, 612 00 Brno, Czech Republic}

\author{Olga Fedotova}
%\email{klapetek@cmi.cz}
     \affiliation{Scientific and Practical Materials Research Center, National Academy of Sciences of Belarus,  P. Brovki str. 19, 220072, Minsk,  Belarus}

\begin{abstract}
%
%ABSTRACT TO BE FILLED PK: AUTHOR ORDERING PRELIMINARY I HAVE JUST PUT IT THERE SO THE SPACE IS RESERVED FOR THE TEXT LENGTH..
%

%ABSTRACT TO BE ADDED!

We study the effect of Coulomb correlation on the emission properties of the ground state exciton in zincblende CdSe/ZnS core-shell and in wurtzite ZnO quantum dots. We validate our theory model by comparing results of computed exciton energies of CdSe/ZnS quantum dots to photoluminescence and scanning near-field optical microscopy measurements. We use that to estimate the diameter of the quantum dots using a simple model based on infinitely deep quantum well and compare the results with the statistics of the atomic force microscopy scans of CdSe/ZnS dots, obtaining excellent agreement. Thereafter, we compute the energy fine structure of exciton, finding striking difference between properties of zincblende CdSe/ZnS and wurtzite ZnO dots. While in the former the fine structure is dominated by the dipole terms of the multipole expansion of the exchange interaction, in the latter system that is mostly influenced by Coulomb correlation. Furthermore, the correlation sizeably influences also the exciton binding energy and emission radiative rate in~ZnO dots. 
%
%The~reason for that is found to be the topology of the hole wavefunction caused by the symmetry breaking in the wurtzite crystal structure.

%
\end{abstract}

\maketitle

\iffalse

{\bf
WHAT STILL NEEDS TO BE ADDED:

\begin{itemize}
    \item Introduction into II-VI QD system including references (that will be pain),
    \item Graphs of measurements of PL, SNOM, and AFM of CdSe QDs,
    \item Reasonably short/long description of experiment+citations on that,
    \item Calculations of light propagation from ZnO QDs at the end of the paper.
    \item ...
\end{itemize}
}

\fi

\section{Introduction}

% Here some intro about CdSe/ZnS and ZnO QDs should be given along with references

Semiconductor quantum dots (QDs) grown by epitaxy from materials belonging to group III and V of the Periodic table are~one of the most promising quantum light source in quantum technology, as they combine excellent optical properties with the compatibility to semiconductor processing and the potential for scalability.~\cite{Aharonovich2016,Senellart2017,Thomas2021,Tomm2021,Orieux2017,Huber2018a,Klenovsky_IOP2010,Klenovsky2015,Klenovsky2012,Sala2018} Meanwhile, they provide also a platform for photon-to-spin conversion~\cite{Atature2018,Borri2001}, building up bridges between photonic and spin qubits.~\cite{Gangloff2019,Chekhovich2020,Krapek2010} Moreover, they can~be~used as~building blocks for quantum information devices, particularly for quantum repeaters~\cite{Bimberg2008_EL,Azuma_Qrep,Li2019}, as efficient single and entangled photon sources~\cite{Lochamnn2006,muller_quantum_2018,martin-sanchez_single_2009, schlehahn_single-photon_2015,paul_single-photon_2017,salter_entangled-light-emitting_2010,Aberl2017,Klenovsky2018, Senellart2017,Csontosova2020,Huber2019}, including highly-entangled  states for quantum computing~\cite{Lim_PRL2005,Lindner_PRL2009,Istrati2020,Klenovsky2016,Steindl2021}, or as nanomemories.~\cite{Marent2011,BimbergPatent,Marent2009_microelectronics,Bimberg2011_SbQDFlash, Marent_APL2007_10y,Sala2018,Klenovsky2019} One of the drawbacks of the aforementioned QD technologies is their elevated cost caused by the epitaxial grown techniques used for their fabrications,~i.e., Metal-Organic Vapor Phase Epitaxy (MOVPE) or Molecular Beam Epitaxy (MBE).

Compared to the epitaxially grown QDs there exist another class of quantum dots based on solution processed semiconducting nanocrystals with dimensions smaller than $\sim$\,20\,nm, so called colloidal QDs.~\cite{Liu2021,Park2021} Since their emergence more than 20 years ago~\cite{Ekimov1985,Rossetti1998} the ease of~their manufacture and relatively low cost enabled their use in many optoelectronic applications such as~lasers sources~\cite{Park2021}, light-emitting diodes (LEDs)~\cite{Coe2002,Caruge2008}, photodetectors~\cite{Konstantatos2006}, and solar cells.~\cite{Nozik2002,Pattantyus-Abraham2010} Moreover, similarly to epitaxially grown QDs, they were utilized in~integrated-photonic circuits~\cite{Xie2017,Cegielski2018,Kim2018,Amemiya2017}, lab-on-chip platforms~\cite{Vannahme2011}, optical interconnects~\cite{Jung:17,Clark2010,Grivas2012}, or advanced medical devices.~\cite{Vollmer2008,Chandrasekar2019,Retolaza2016,Chen2019} Moreover, colloidal QDs were realized for different material systems from the Periodic table groups II-VI, III-V, or IV-VI. One of their advantages is that the properties of their quantum confinement depend predominantly on their diameter which can be~controlled by the growth. 

In this work we investigate the electronic structure of~CdSe/ZnS~\cite{Raino2012,Moreels2011} core-shell and ZnO~\cite{Fonoberov2006,Fonoberov2004,Rahman2011,Li2011} QDs. We~consider the crystal structure of the former system to be that of the zincblende crystal while the latter is~considered in the wurtzite phase. We focus here on the~properties of the ground state exciton ($X^0$), in particular investigating energy, fine-structure splitting (FSS), energy difference between dark and bright (BD) states and emission rate of $X^0$. We compute the uncorrelated single-particle (SP) states of electrons and holes using the envelope function approximation based on eight-band ${\bf k\cdot p}$ theory, the results of which are used thereafter as basis states for the configuration interaction (CI) computations which elucidate the Coulomb interaction energies and correlation. We first test our theory model by comparing the results for CdSe/ZnS QDs with photoluminescence (PL) spectroscopy, scanning near-field optical microscopy (SNOM), and time-resolved PL (TRPL). Thereafter, we compare the aforementioned theory results to that obtained on ZnO QDs.

\section{Methods}

\subsection{Single-particle states \& configuration interaction}

% About k.p and CI (copied and slightly changed from appendix of PRB 104, 165401 (2021) with AR and Hui because there it is described the best way so far.

We first give a description of our method of computation.~\cite{Csontosova2020} We start with obtaining SP states using the~envelope function method based on eight-band $\mathbf{k}\cdot \mathbf{p}$ approximation using the Nextnano++ simulation suite.~\cite{Birner2007} SP states obtained that way read
\begin{equation}
    \Psi_{a_i}(\mathbf{r}) = \sum_{\nu\in\{s,x,y,z\}\otimes \{\uparrow,\downarrow\}} \chi_{a_i,\nu}(\mathbf{r})u^{\Gamma}_{\nu}\,,
\end{equation}
where $u^{\Gamma}_{\nu}$ is the Bloch wave-function of an $s$-like conduction band or a $p$-like valence band at the center of~the~Brillouin zone ($\Gamma$ point), $\uparrow$/$\downarrow$ mark the spin, and $\chi_{a_i,\nu}$ is~the~envelope function, where $a_i \in \{ e_i, h_i \}$.

Those states are then used as basis states for the CI method~\cite{SFZ01,Schliwa:09} using the code which we have previously developed.~\cite{Klenovsky2017,Klenovsky2018_TUB,Csontosova2020} Let us assume that the excitonic complex $\ket{\rm M}$ consists of $N_e$ electrons and $N_h$ holes. The~CI method uses as a basis the Slater determinants (SDs) consisting of $n_e$ SP electron and $n_h$ SP hole states. 

The trial function of the excitonic complex then reads
\begin{equation}
    \ket{\rm M} =  \sum_{\mathit m=1}^{n_{\rm SD}} \mathit \eta_m \ket{D_m^{\rm M}}, \label{eq9}
\end{equation}
where $n_{\rm SD}$ is the number of SDs in $\ket{D_m^{\rm M}}$, and $\eta_m$ is the constant that is solved for using the variational method. The $m$-th SD can be found as
\begin{equation}
\ket{D_m^{\rm M}} = \frac{1}{\sqrt{N!}} \sum_{\tau \in S_N} \rm sgn \mathit(\tau) \phi_{\tau\{i_1\}}(\mathbf{r}_1) \phi_{\tau\{i_2\}}(\mathbf{r}_2) \dots \phi_{\tau\{i_N\}}(\mathbf{r}_N).
\end{equation}
Here, we sum over all permutations of $N~:=~N_e~+~N_h$ elements over the symmetric group $S_N$. For the sake of notational convenience, we joined the electron and hole wave functions of which the SD is~composed of, in a unique set $\{\phi_1, \dots, \phi_N\}_m := \{ \Psi_{e_j}, \dots,\Psi_{e_{j+N_e-1}}; \Psi_{h_k}, \dots,\Psi_{h_{k+N_h-1}} \}$, where $j \in \{1,\dots, n_e \}$ and $k \in \{1,\dots, n_h \}$. Accordingly, we join the positional vectors of electrons and holes $\{r_1, \dots, r_N\}:= \{ \mathbf{r}_{e_1}, \dots,\mathbf{r}_{e_{N_e}}; \mathbf{r}_{h_1}, \dots,\mathbf{r}_{h_{N_h}} \}$.

Thereafter, we solve within our CI the Schr\"{o}dinger equation 
\begin{equation}
\hat{H}^{\rm{M}} \ket{\rm{M}} = E^{\rm{M}} \ket{\rm{M}} ,
\end{equation}
where $E^{\rm{M}}$ is the eigenenergy of excitonic state $\ket{\rm{M}}$, and~$\hat{H}^{\rm{M}}$ is the CI Hamiltonian which reads $\hat{H}^{\rm{M}}~=~\hat{H}_0^{\rm{M}}~+~\hat{V}^{\rm{M}}$, where $\hat{H}_0^M$ represents the SP Hamiltonian and $\hat{V}^{\rm{M}}$ is~the~Coulomb interaction between SP states. The matrix element of $\hat{V}^{\rm{M}}$ reads~\cite{Klenovsky2017,Klenovsky2019,Csontosova2020}
\begin{equation}
\begin{split}
    &\bra{D_n^{\rm M}}\hat{V}^{\rm{M}}\ket{D_m^{\rm M}} = \frac{1}{4\pi\epsilon_0} \sum_{ijkl} \iint {\rm d}\mathbf{r} {\rm d}\mathbf{r}^{\prime} \frac{q_iq_j}{\epsilon(\mathbf{r},\mathbf{r}^{\prime})|\mathbf{r}-\mathbf{r}^{\prime}|} \\
    &\times \{ \Psi^*_i(\mathbf{r})\Psi^*_j(\mathbf{r}^{\prime})\Psi_k(\mathbf{r})\Psi_l(\mathbf{r}^{\prime}) - \Psi^*_i(\mathbf{r})\Psi^*_j(\mathbf{r}^{\prime})\Psi_l(\mathbf{r})\Psi_k(\mathbf{r}^{\prime})\}.
\end{split}
\label{eq:CoulombMatrElem}
\end{equation}
In Eq.~\eqref{eq:CoulombMatrElem} $q_i$ and $q_j$ label the elementary charge $|e|$ of either electron ($-e$), or hole ($e$), and $\epsilon(\mathbf{r},\mathbf{r}^{\prime})$ is the spatially dependent dielectric function. Note, that the Coulomb interaction is treated as a perturbation. The evaluation of the sixfold integral in Eq.~\eqref{eq:CoulombMatrElem} is performed using the~Green's function method~\cite{Schliwa:09,Stier2000,Klenovsky2017}
\begin{equation}
\begin{split}
    \nabla \left[ \epsilon(\mathbf{r}) \nabla \hat{U}_{ajl}(\mathbf{r}) \right] &= \frac{4\pi e^2}{\epsilon_0}\Psi^*_{aj}(\mathbf{r})\Psi_{al}(\mathbf{r}),\\
    V_{ij,kl} &= \int{\rm d}\mathbf{r}^{\prime}\,\hat{U}_{ajl}(\mathbf{r}^{\prime})\Psi^*_{bi}(\mathbf{r}^{\prime})\Psi_{bk}(\mathbf{r}^{\prime})\,,
\end{split}
\end{equation}
where $a,b \in \{e,h\}$ and $\nabla := \left( \frac{\partial}{\partial x}, \frac{\partial}{\partial y}, \frac{\partial}{\partial z} \right)^T$. Note that $\epsilon(\mathbf{r},\mathbf{r}^{\prime})$ was set to bulk values~\cite{Klenovsky2019,Csontosova2020} for the CI calculations presented here. 

We note that in this work we consider only excitons,~i.e., $N_e=1$ and $N_h=1$, respectively. However, in~order to capture the effect of Coulomb correlation we~vary the number of SP basis states $n_e$ and $n_h$ from 2~to~12. 

Finally, we note that the multipole expansion of the exchange interaction was included in our CI for SP basis of $n_e=2$ and $n_h=2$ following the theory outlined in Refs.~\cite{Takagahara2000,Krapek2015}.

\subsection{Atomic force microscopy}
Atomic force microscopy~\cite{Klapetek2018} (AFM) of CdSe QDs on Si substrate was measured by the Bruker Dimension ICON AFM with Bruker RTESPA-300 probe. We used the program Gwyddion~\cite{Necas2012} to analyze the AFM topographic images. Since the microscope probe has typically larger radius than the size of the QD, the lateral dimensions of measured QDs are highly affected by tip convolution, which leads to their effective broadening. Therefore the size distribution was calculated on basis of~QDs maxima, using the flat sample substrate as a reference.

\subsection{Photoluminescence spectroscopy}

Standard PL setup was used in PL and TRPL measurements. In both PL and TRPL the excitation of the sample was done by a pulsed laser (510\,nm, 30\,MHz) with emission intensity varied by a neutral density (ND) filter from 0 to $400\,\mu$W. The samples were positioned in~a~cryostat, cooled to 70\,K and were heated up to room temperature. The PL spectrum was detected by an Andor CCD camera in a visible part of a spectrum and the TRPL signal was measured by an avalanche diode connected to~the~time triggered photon counting device.

\subsection{Scanning near-field optical microscopy}

High spatial resolution PL spectra were obtained using Scanning near-field optical microscopy (SNOM) setup. PL spectra were measured using commercial Thermomicroscopes Aurora II aperture SNOM enhanced for spectroscopic measurements. Standard SNOM probes with 100\,nm aperture manufactured by Thermomicroscopes were used. The CdSe/ZnS QDs on Si substrate were excited by an Ar ion laser (488\,nm) and PL spectra were acquired using Avantes AvaSpec HS-TEC spectrometer at room temperature. The integration time was 60\,s. In~order to~increase signal-to-noise ratio, we measured two spectra, one with excitation laser on, and another spectrum with excitation laser off. The SNOM PL spectrum shown in Fig.~\ref{fig:CdSe_ExpTeor}~(d) was obtained as a fraction of these two datasets.
%normalized to the PL spectra measured at 70~K.

\section{Results}

\subsection{CdSe/ZnS zincblende quantum dots}

We start our investigation of II-VI QDs with the dicussion of QDs consisting of CdSe core and ZnS shell. We have modeled CdSe/ZnS QDs as spheres and defined the structure using Nextnano++ simulation suite.~\cite{Birner2007} Since one of the most important properties of those dots is their diameter, we modulated that in our calculations discussed in the following. Furthermore, motivated by typical properties of commercially available CdSe/ZnS dots, we have set the thickness of the ZnS shell to 1\,nm

\onecolumngrid

\begin{figure} [!ht]
	\centering
	\includegraphics[scale=0.83]{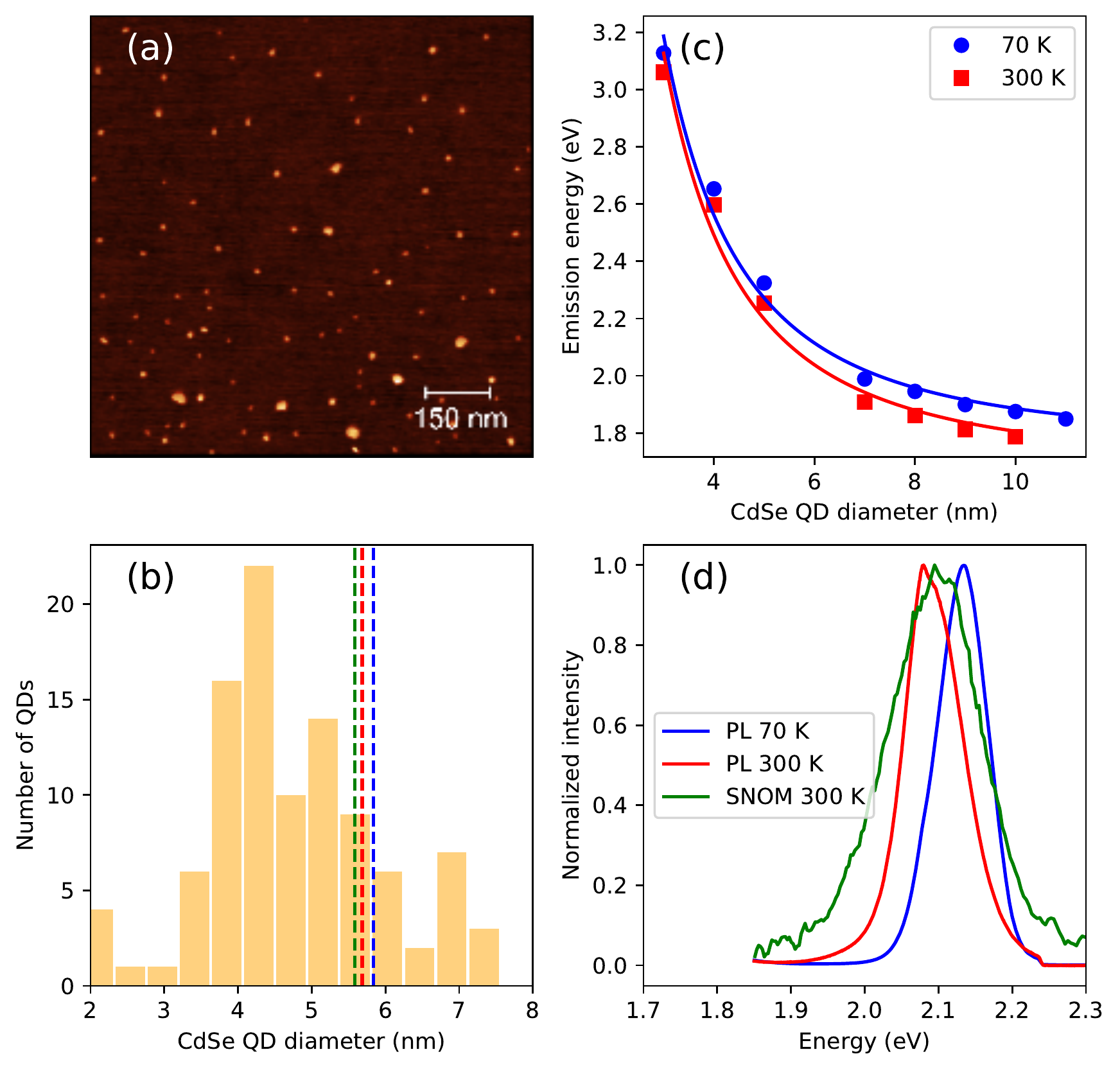}
	\caption{Comparison of the experiment and theory of CdSe/ZnS QDs. In panel (a) we show an AFM scan of commercially available CdSe/ZnS QDs emitting in yellow part of the spectrum. Panel (b) gives the statistics of QD diameters from (a) obtained using Gwyddion~\cite{Necas2012} program as yellow bars. In (c) are the computed emission energies (symbols) of $X^0$ as a function of dot diameter for two sample temperatures. The calculations were performed by eight-band ${\bf k.p}$~\cite{Birner2007} and CI, the latter with full inclusion of the Coulomb correlation~\cite{Klenovsky2017,Klenovsky2018_TUB,Csontosova2020}. The full curves in (c) represent the fits of the theory data by our simplified model given in Eq.~\eqref{eq:infiniteQW} (see also text). The PL and SNOM measurements of CdSe/ZnS QDs are shown in panel (d). Note that the broken lines in~(b) correspond to energy positions of PL and SNOM maxima in (d) recomputed to QD diameter using fitted data from (c).}
	\label{fig:CdSe_ExpTeor}
\end{figure}

\twocolumngrid

\noindent in all our calculations. Since our QDs are not embedded in any bulk material nor grown by epitaxy, we do not minimize the elastic energy in the whole simulated structure,~i.e., we consider our dots to be strain free. That means, we assume that the ZnS shell around the CdSe dot does not lead to any induced strain in the dot's CdSe core. That assumption is reasonable, since capping with a 1\,nm layer of material seems to be unlikely to influence the much more bulky CdSe core of the dot. Finally the~eigenenergies and wavefunctions resulting from SP calculations are fed into the CI solver and we obtain states of~correlated excitons as described above.

% HERE A FIGURE OF PL/SNOM/AFM OF CDSE QDS INCLUDING THE CORRESPONDING RESULTS OF CALCULATION FOR 300K SHOULD BE PUT

\iffalse
\onecolumngrid

\begin{figure} [!ht]
	\centering
	%\includegraphics[]{./figure/figure1}
	\includegraphics[scale=0.8]{CdSe_QD_AFM_stat_E_PL_300K.pdf}
	\caption{Comparison of the experiment and theory of CdSe/ZnS QDs. In panel (a) we show an AFM scan of commercially available CdSe/ZnS QDs emitting in yellow part of the spectrum. Panel (b) gives the statistics of QD diameters from (a) obtained using Gwyddion~\cite{Necas2012} program as yellow bars. In (c) are the computed emission energies (symbols) of $X^0$ as a function of dot diameter for two sample temperatures. The calculations were performed by eight-band ${\bf k.p}$~\cite{Birner2007} and CI, the latter with full inclusion of the Coulomb correlation~\cite{Klenovsky2017,Klenovsky2018_TUB,Csontosova2020}. The full curves in (c) represent the fits of the theory data by our simplified model given in Eq.~\eqref{eq:infiniteQW} (see also text). The PL and SNOM measurements of CdSe/ZnS QDs are shown in panel (d). Note that the broken lines in~(b) correspond to energy positions of PL and SNOM maxima in (d) recomputed to QD diameter using fitted data from (c).}
	\label{fig:CdSe_ExpTeor}
\end{figure}

\twocolumngrid
\fi
% HERE A FIGURE OF PL/SNOM/AFM OF CDSE QDS INCLUDING THE CORRESPONDING RESULTS OF CALCULATION FOR 300K SHOULD BE PUT

%\newpage

However, prior to discussing our theory results, we~test the setting of our simulated structure and theory tools described above by comparing that with the results of~measurements using AFM, PL spectroscopy, and SNOM. The corresponding results are given in Fig.~\ref{fig:CdSe_ExpTeor}. A~commercially available sample of CdSe/ZnS QDs emitting in yellow part of the spectra was first measured using AFM \{Fig.~\ref{fig:CdSe_ExpTeor}~(a)\} and the results of that were statistically analyzed using the Gwyddion~\cite{Necas2012} program \{Fig.~\ref{fig:CdSe_ExpTeor}~(b)\}. We~focused particularly on the distribution of QD's diameters on the sample in order to compare to our calculations shown by symbols in Fig.~\ref{fig:CdSe_ExpTeor}~(c) for sample temperatures of 70 and 300\,K. The calculations were performed by~eight-band ${\bf k.p}$~\cite{Birner2007} and CI, the latter with full inclusion of the Coulomb correlation~\cite{Klenovsky2017,Klenovsky2018_TUB,Csontosova2020}. For better transferability of the theory model and to ease comparison with experiment, we have fitted that data using a~simple model motivated by results for ground state of~the~infinitely deep quantum well and given by
\begin{equation}
\label{eq:infiniteQW}
E=\frac{a}{D^2}+E_{\infty}\,,
\end{equation}
where $D$ marks the QD diameter, $E_{\infty}$ the energy for $D\rightarrow \infty$, and an additional fitting parameter $a$. Note that the latter two parameters ($E_{\infty}$ and $a$) neatly lump the material properties of our QDs. Finally, we have measured emission from our CdSe/ZnS QDs using PL and SNOM \{Fig.~\ref{fig:CdSe_ExpTeor}~(d)\}. Furthermore, we have picked the energies $E_{PL}$ corresponding to maxima in Fig.~\ref{fig:CdSe_ExpTeor}~(d) and recomputed that to QD diameter using Eq.~\eqref{eq:infiniteQW} as $D=\sqrt{a/(E_{PL}-E_{\infty})}$, where $a$ and $E_{\infty}$ were obtained from fitting in Fig.~\ref{fig:CdSe_ExpTeor}~(c). The resulting QD diameters are~shown by broken curves in Fig.~\ref{fig:CdSe_ExpTeor}~(b) and are satisfactorily close to mean QD diameter of 5.5\,nm observed by AFM. 

\iffalse
\onecolumngrid

\begin{figure} 
	\centering
	%\includegraphics[]{./figure/figure1}
	\includegraphics[scale=0.9]{CdSe_QD_AFM_stat_E_PL_300K.pdf}
	\caption{Comparison of the experiment and theory of CdSe/ZnS QDs. In panel (a) we show an AFM scan of commercially available CdSe/ZnS QDs emitting in yellow part of the spectrum. Panel (b) gives the statistics of QD diameters from (a) obtained using Gwyddion~\cite{Necas2012} program as yellow bars. In (c) are the computed emission energies (symbols) of $X^0$ as a function of dot diameter for two sample temperatures. The calculations were performed by eight-band ${\bf k.p}$~\cite{Birner2007} and CI, the latter with full inclusion of the Coulomb correlation~\cite{Klenovsky2017,Klenovsky2018_TUB,Csontosova2020}. The full curves in (c) represent the fits of the theory data by our simplified model (see text). The PL and SNOM measurements of CdSe/ZnS QDs are shown in panel (d). Note that the broken lines in (b) correspond to energy positions of PL and SNOM maxima in (d) recomputed to QD diameter using fitted data from (c).}
	\label{fig:CdSe_ExpTeor}
\end{figure}

\twocolumngrid
\fi

\begin{figure}
	\centering
	\includegraphics[scale=0.8]{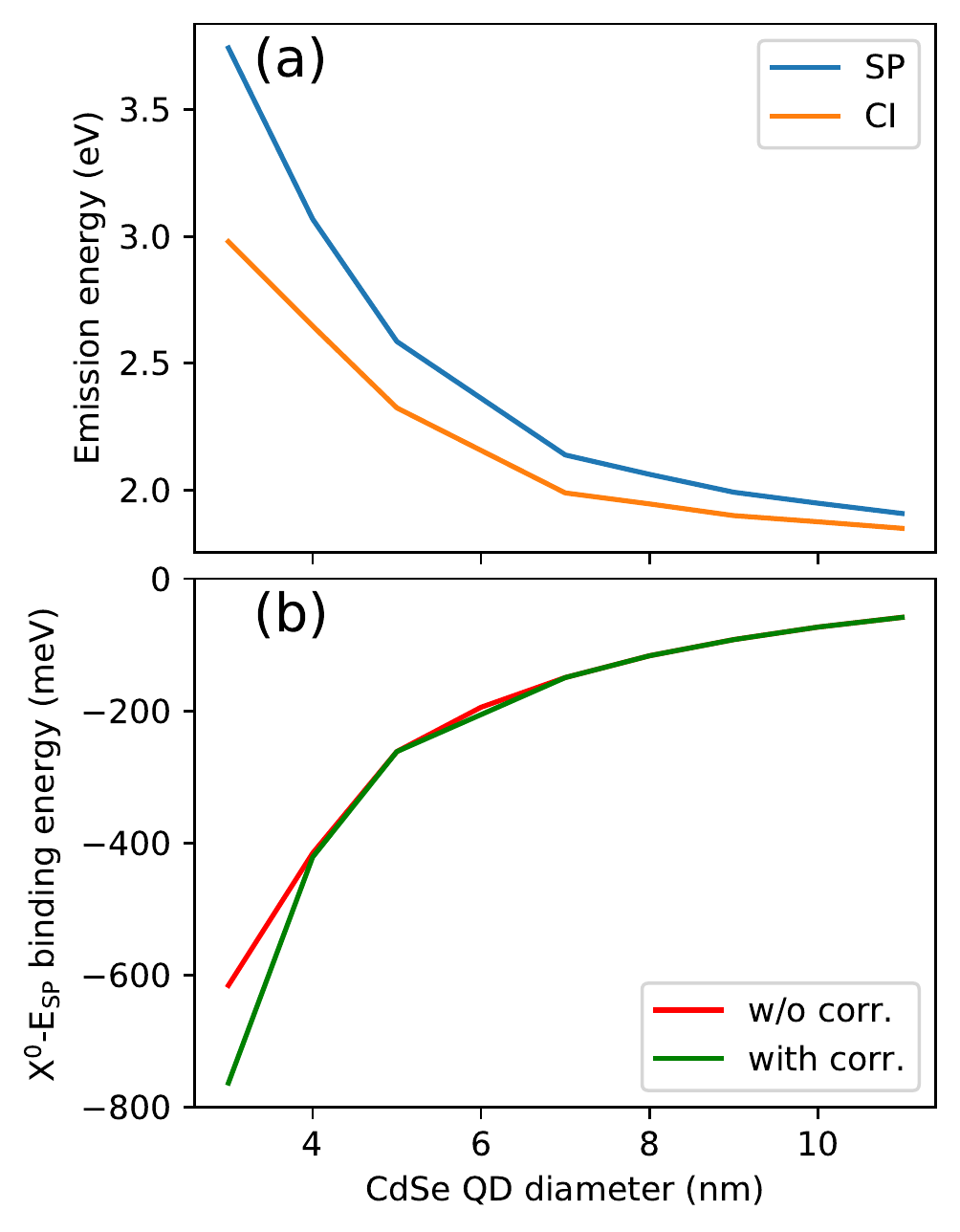}
	\caption{Energy difference between ground state electron and hole single particle (SP) states and the corresponding energy of the exciton computed by the configuration interaction~\cite{Klenovsky2017,Klenovsky2019,Csontosova2020} (CI) method as a function of CdSe QD diameter are shown in panel (a). Notice the difference between results for SP and CI computations, being due to direct Coulomb interaction. Binding energy of the exciton computed by CI with the SP basis of 2~electron and 2~hole SP ground states (w/o corr.) and that for 12 electron and 12 hole SP states (with corr.) as a function of ZnO QD diameter is given in panel (b). Notice that the effect of the Coulomb correlation is found to be negligible for binding energy of $X^0$ in CdSe QDs. The calculations were performed for temperature of 70\,K.
	%Notice the difference between results for SP and CI computations, being due to direct Coulomb interaction further extracerbated by CdSe dielectric constant of $\epsilon=9.7$~\cite{landoltbornstein}.
	}
	\label{fig:CdSe_EB}
\end{figure}

Thus, having validated the correct settings of our simulation toolbox, we proceed with discussion of the theory results. The SP ground state electron-hole transition energy as well as that for $X^0$ for temperature of 70\,K for CdSe/ZnS QDs is given in Fig.~\ref{fig:CdSe_EB}. We first notice in~Fig.~\eqref{fig:CdSe_EB}~(a) the sizeable difference between the results obtained using SP and CI methods, clearly showing that SP computations only give quite imprecise results for these dots. The difference between SP and CI results originates from the attractive direct electron-hole Coulomb integral $J_{eh}$ considered in CI. 

The impact of $J_{eh}$ can be viewed in Fig.~\ref{fig:CdSe_EB}~(b) in the form of $X^0$ binding energy with respect to the corresponding SP result ($E_{sp}$). Moreover, in that panel we compare also the results of the binding energy obtained using CI with basis of two electron and two hole SP basis states (2x2) and that obtained using 12 electron and 12~hole SP states (12x12), respectively. While the CI result with 2x2 basis includes a minimum of Coulomb correlation, that for 12x12 contains almost the full effect of the Coulomb correlation. Clearly, correlation has negligible effect on binding energy of $X^0$ in CdSe/ZnS QDs. Notice that compared to similar results in III-V QDs~\cite{Klenovsky2017} the magnitude of $J_{eh}$ in CdSe/ZnS QDs is~further extracerbated by the dielectric constant of CdSe of~$\epsilon=9.7$~\cite{landoltbornstein} being almost two times smaller compared to,~e.g., InAs ($\epsilon=15.15$~\cite{landoltbornstein}) or GaAs ($\epsilon=12.93$~\cite{Birner2007}), what effectively increases the magnitude of the Coulomb integrals in CI since $\epsilon$ is in the denominator of that, see Eq.~\eqref{eq:CoulombMatrElem}.

\begin{figure}
	\centering
	\includegraphics[scale=0.8]{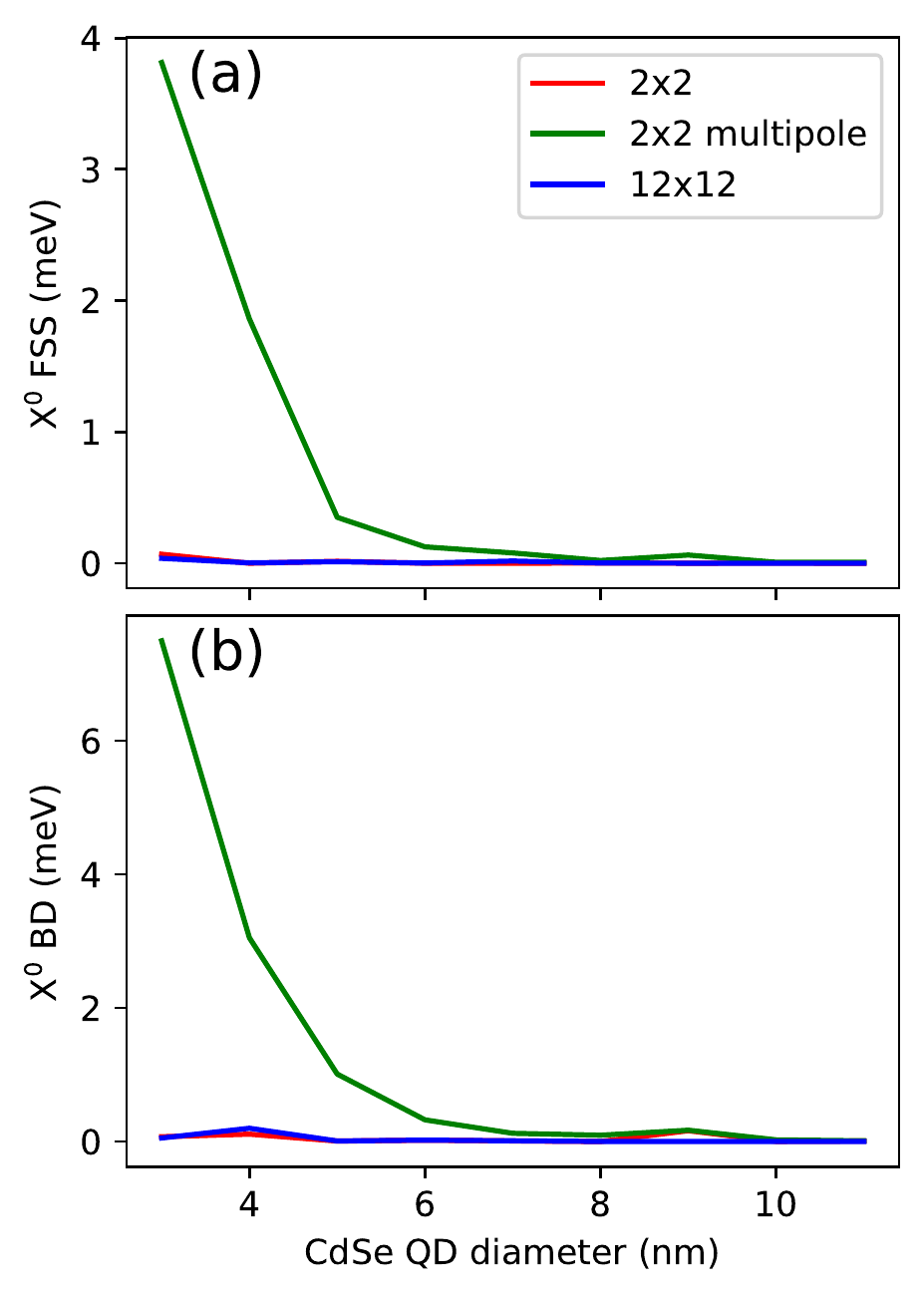}
	\caption{Panel (a) gives the CdSe QD diameter dependence of the fine structure splitting ($X^0$ FSS) of~the~bright exciton computed with the SP basis of~2~electron and 2~hole SP ground states (2x2)~\cite{Klenovsky2017}, that including the multipole expansion of~the~exchange interaction~\cite{Krapek2015} (2x2 multipole), and computation for CI basis of 12 electron and 12 hole SP states (12x12). In panel (b) the energy difference between bright and dark excitonic doublet ($X^0$ BD) as a function of ZnO QD diameter is shown. Notice that both FSS and BD are dominated by multipole expansion terms of the exchange interaction in~CdSe QDs. The calculations were performed for temperature of 70\,K.}
	\label{fig:CdSe_FSS}
\end{figure}

Next, we study the energy splitting of the bright exciton Kramers doublet of $X^0$ of CdSe/ZnS QDs,~i.e., FSS, and the energy difference between the optically bright and dark doublet of $X^0$,~i.e., BD. The results are given in Fig.~\ref{fig:CdSe_FSS}. Here, we study the effect of various physical phenomena influencing the exchange interaction and, thus, FSS and BD. Namely, those are (i) the exchange Coulomb interaction between ground state electron and hole wavefunction (marked as ``2x2" in Fig.~\ref{fig:CdSe_FSS}), (ii) the multipole expansion of exchange interaction~\cite{Krapek2015} (marked as ``2x2 multipole" in Fig.~\ref{fig:CdSe_FSS}), and (iii) the effect of Coulomb correlation~\cite{Klenovsky2017,Huang2021} (marked as ``12x12" in Fig.~\ref{fig:CdSe_FSS}). Clearly, the dominant contribution to both FSS and BD in CdSe/ZnS QDs comes from the multipole expansion of the exchange interaction, similarly as~in~Refs.~\cite{Krapek2015,Krapek2016} for III-V QDs. We further note, that the dominant contribution comes from the dipole-dipole term of that expansion for CdSe QDs. We finally note that our results quantitatively agree with previous experiments~\cite{Htoon2008} as well as atomistic theoretical results~\cite{Bui2020}.

\begin{figure}
	\centering
	\includegraphics[scale=0.8]{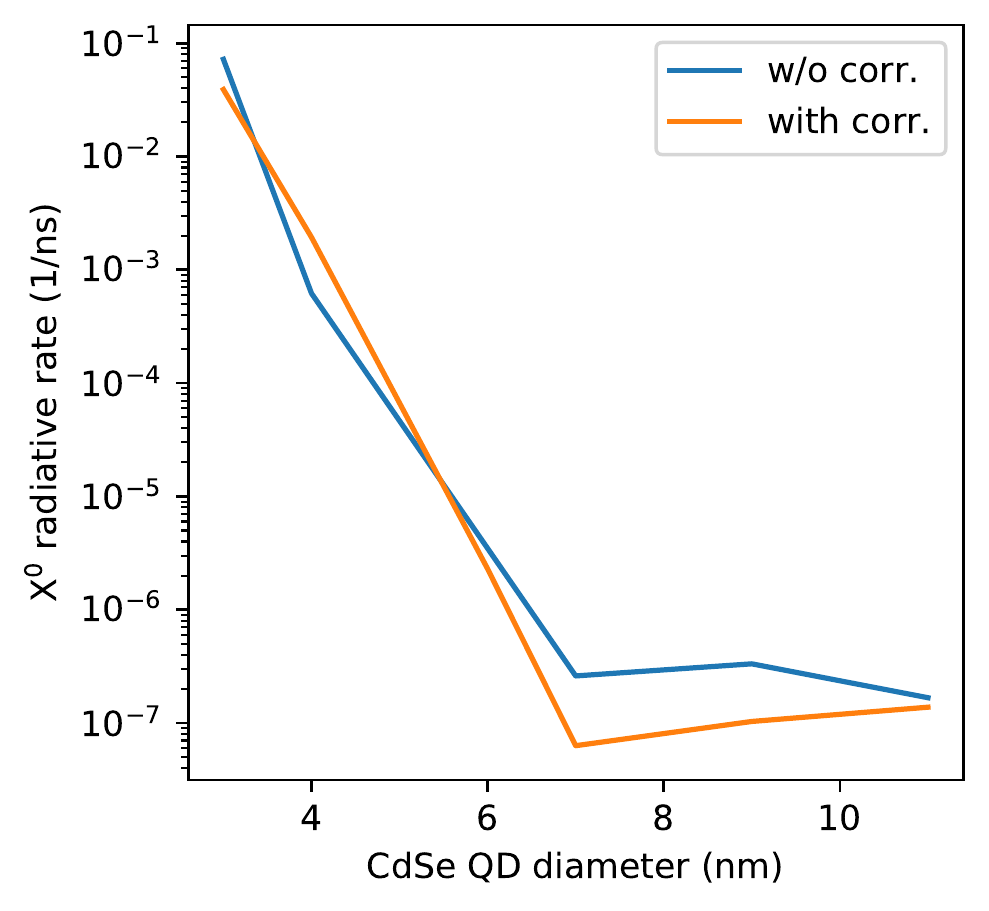}
	\caption{Radiative rate of bright exciton as a function of CdSe QD diameter. The data are shown for calculation using CI with basis of 2~electron and 2~hole SP ground states (w/o corr.) and that for 12 electron and 12 hole SP states (with corr.). Notice that Coulomb correlation has negligible effect on the emission of CdSe QDs. The calculations were performed for temperature of 70\,K.}
	\label{fig:CdSe_rate}
\end{figure}

Furthermore, alongside the CI calculations we have computed also the emission radiative rate~\cite{Klenovsky2017} of the recombination of $X^0$ for CdSe/ZnS QDs using the Fermi's golden rule~\cite{Dirac1927}, see Fig.~\ref{fig:CdSe_rate}. As can be seen, the computations of the rate with and without the inclusion of Coulomb correlation lead to similar results,~i.e., the dominant contribution to that comes already from the recombination of the electron and hole ground states. Furthermore, we notice that the rate reduces considerably with QD size. However, since we compute the rate using the Fermi's golden rule between the envelope functions only, we admit that our computed results might provide smaller values than those seen in experiment. Nevertheless, the~obtained values are close to those experimentally obtained from time-resolved photoluminescence measurements (TRPL).

%We have measured TRPL of CdSe/ZnS emitting at a yellow part of the spectrum.
We have tested that by measuring TRPL of an ensemble of our CdSe/ZnS QDs. 
%We counted single photons and then measured a decay of the intensity after a laser pulse. 
The reduction of the single photon rate after each laser pulse was fitted by a double mono-exponential (2ME) decay model, see also Ref.~\cite{Steindl2018}
\begin{equation}
\label{qe:TRPLmeas}
I(t) = A_1 \exp\{-(t-t_0)/\tau_1\}+A_2 \exp\{-(t-t_0)/\tau_2\},
\end{equation}
where the amplitude of ME $A_1$ ($A_2$) and the decay time $\tau_1$ ($\tau_2$) characterize the slow (fast) decay process and $t_0$ is the time of the QD emission triggered by the pulse laser and the start of the decay of intensity. Note that $A_1+A_2$ is the maximum intensity emitted from QDs. We can see in Fig.~\ref{fig:CdSe,ZnS_TRPL_yellow}~(a) that the next pulse of the laser comes sooner than the slow ME completely decays which we call repumping, the effect of which we included in the fitting.

In order to compare with our theory we introduce $\tau_{PL}$ that represents the weighted arithmetic mean of two decay times $\tau_1$ and $\tau_2$ for 2ME model and is given by
\begin{equation}
\label{47}
\tau_{PL} = \sum_{i=1}^2 w_i \tau_i \,; \,\,\,  w_i= \frac{A_i \tau_i}{\sum_{i=1}^2 A_i \tau_i},
\end{equation}
where $w_i$ is the weight defined by parameters of 2ME model.
\begin{figure}
	\centering
	\includegraphics[scale=0.75]{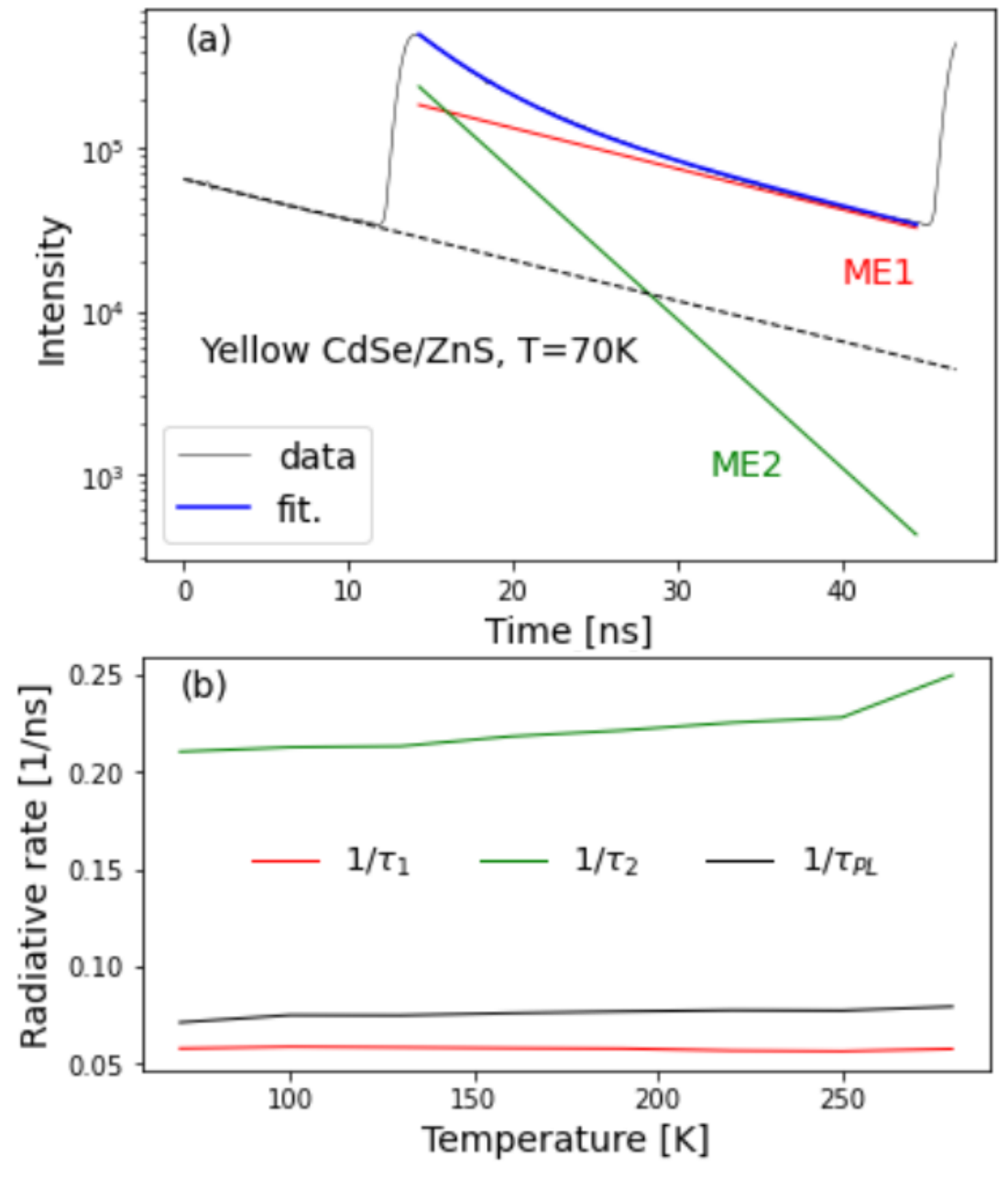}
	\caption{(a) Deconvolved TRPL signal (black solid curve) measured at 70\,K by 2ME model (blue solid curve). The slow (fast) ME is represented by red (green) solid curve and decay intensity of slow ME from a previous pulse by dashed black curve. (b) Fitted radiative rate $1/\tau_1$, $1/\tau_2$ and $1/\tau_{PL}$ as~a~function of temperature.~\cite{Valdhans2021}}
	\label{fig:CdSe,ZnS_TRPL_yellow}
\end{figure}

From Fig.~\ref{fig:CdSe,ZnS_TRPL_yellow}~(b) we can infer that the emission radiative rate of our QDs, which is found almost constant for all studied temperatures, is $\sim$\,0.07\:1/ns, a value close to computations of CdSe QDs in Fig.~\ref{fig:CdSe_rate}.
%
%That comparison with experiment verifies the validity of our modelling procedure.

%\clearpage
%\newpage

\subsection{ZnO wurtzite quantum dots}

\begin{figure}
	\centering
	\includegraphics[scale=0.15]{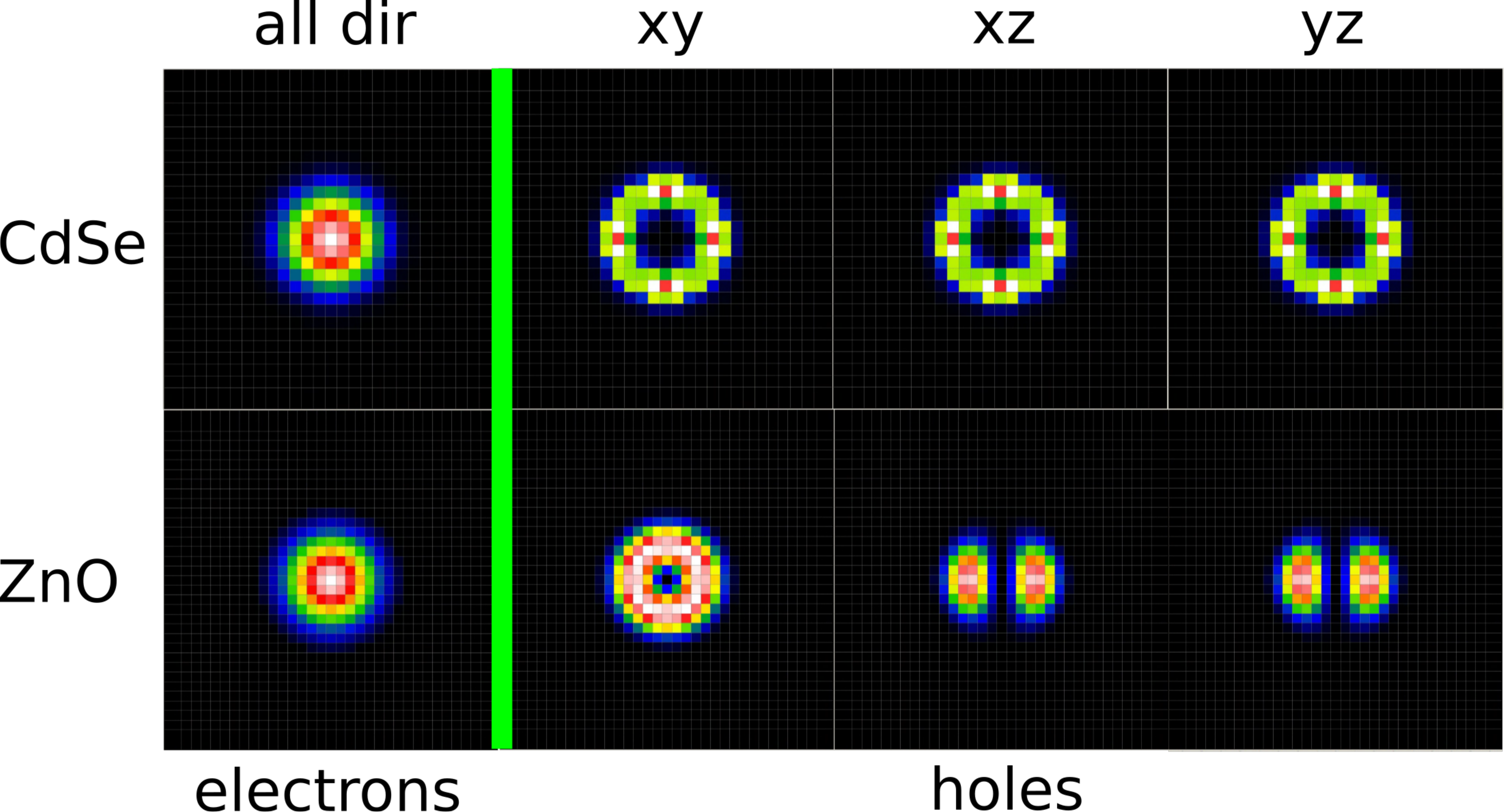}
	\caption{Cuts of through origin of the probability densities of ground state single particle (SP) states of electrons (leftmost column) and holes (remaining three columns) of CdSe/ZnS (top row) and ZnO (bottom row) QDs. The hole densities are shown in three planes defined by the orientation of the corresponding unit vectors ($yz$, $xz$, $xy$). The computations were performed using the Nextnano++ computation suite.~\cite{Birner2007} Notice the difference of the hole probability densities of ZnO QDs between $xy$ plane and that of $xz$ and $yz$ planes originating from the asymmetry of wurtzite crystal along $a$ and $c$ crystal axes (here $c$ is along $z$ axis here). Note that electron ground state SP probability density has spherical symmetry in both studied systems.}
	\label{fig:ZnO_probabden}
\end{figure}

Next II-VI QD system which we study is ZnO wurtzite QDs. Compared to CdSe QDs, the different crystal system in ZnO QDs leads to strikingly different properties of the latter system. The hexagonal crystal structure of~ZnO QDs considerably influences the topology of~the~hole wavefunctions, starting already with the ground state as~can~be~seen in Fig.~\ref{fig:ZnO_probabden}. While for zincblende CdSe QDs, as well as that for III-V QDs~\cite{Klenovsky2010,Klenovsky_IOP2010,Klenovsky2013,Klenovsky2015,Klenovsky2018_TUB,Huang2021}, the hole SP ground state has $s$-like spherical symmetry, in ZnO QDs, the hole ground state has ring-like topology with symmetry axis oriented along $c$-axis of the wurtzite crystal. 

\iffalse
Noticeably, the ring-like topology of the ground state hole SP state causes the reduced overlap with electron ground SP state. Since already the first excited SP hole state has a maximum probability in the center of QD (not shown) and,~thus, increased overlap with ground state electrons, it causes the increase of contribution of the corresponding Coulomb integral in $X^0$. Since the maximum probability density being at the center of ZnO QD is a property also for several other hole excited SP states, that in turn leads to the increase of the importance of Coulomb correlation in $X^0$ for ZnO QD system.
\fi

\begin{figure}
	\centering
	\includegraphics[scale=0.8]{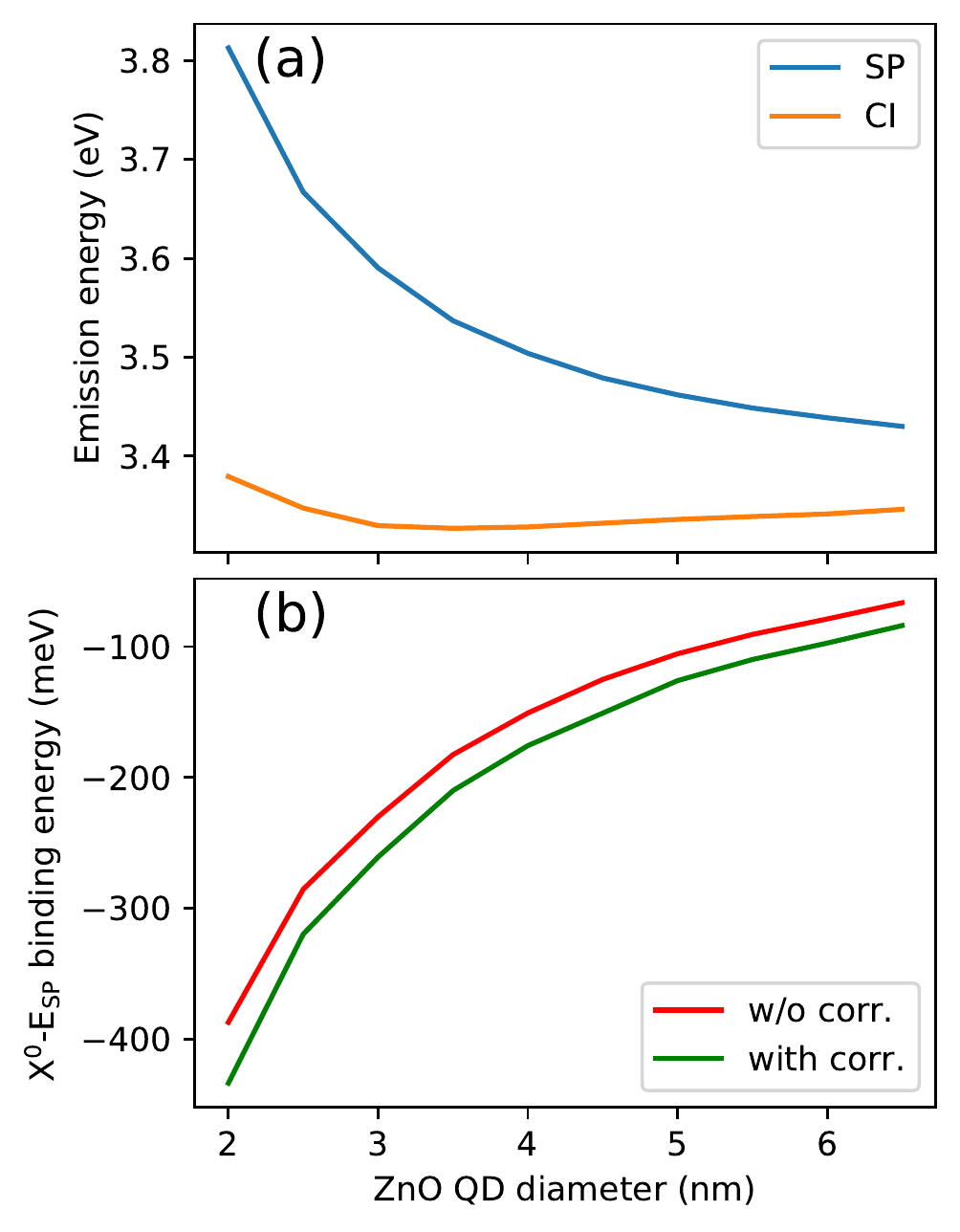}
	\caption{(a) Energy difference between SP and CI states and (b) binding energy of the exciton for ZnO QDs. The outline and marking is the same as that in Fig.~\ref{fig:CdSe_EB}.}
	\label{fig:ZnO_EB}
\end{figure}

%The aforementioned can be seen already in Fig.~\ref{fig:ZnO_EB}~(b), where c
%
Moreover, in ZnO QDs the Coulomb correlation considerably increases the binding energy of $X^0$ with respect to the SP ground state electron-hole transition by $\sim$\,50\,meV as compared to the uncorrelated result. Of course, the dominant contribution to~binding energy seen in Fig.~\ref{fig:ZnO_EB} is due to~the~direct Coulomb integrals $J_{eh}$ similarly as that for CdSe QDs in Fig.~\ref{fig:CdSe_EB} and discussed in the aforementioned.

\begin{figure}
	\centering
	\includegraphics[scale=0.8]{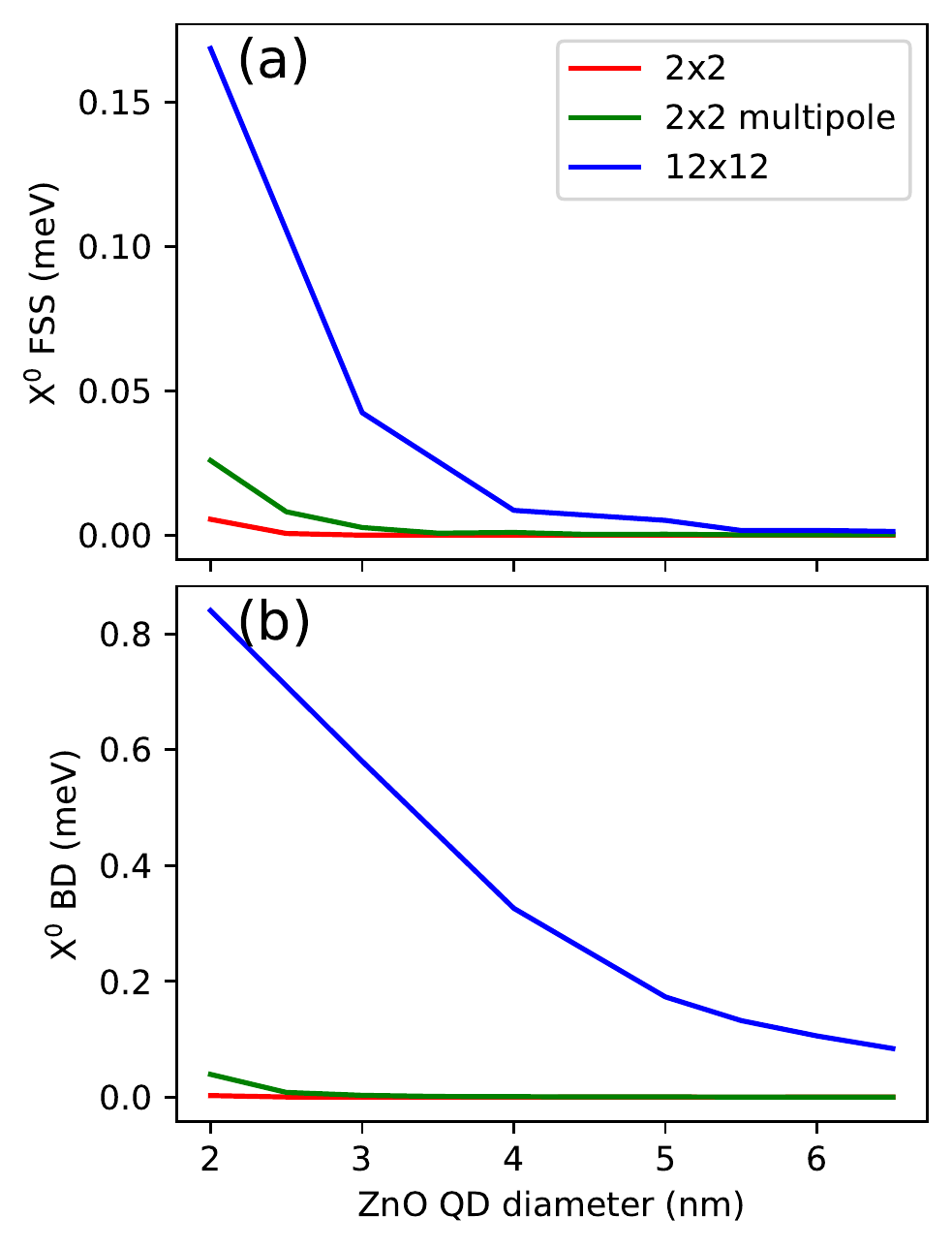}
	\caption{(a) FSS and (b) BD for ZnO QDs as a function of dot diameter. The outline of the figure is the same as that in Fig.~\ref{fig:CdSe_FSS}. Notice that (i) both FSS and BD is relatively smaller and (ii) are dominated by Coulomb correlation in ZnO QDs compared to that for CdSe QDs in Fig.~\ref{fig:CdSe_FSS}.}
	\label{fig:ZnO_FSS}
\end{figure}

\begin{figure}
	\centering
	\includegraphics[scale=0.8]{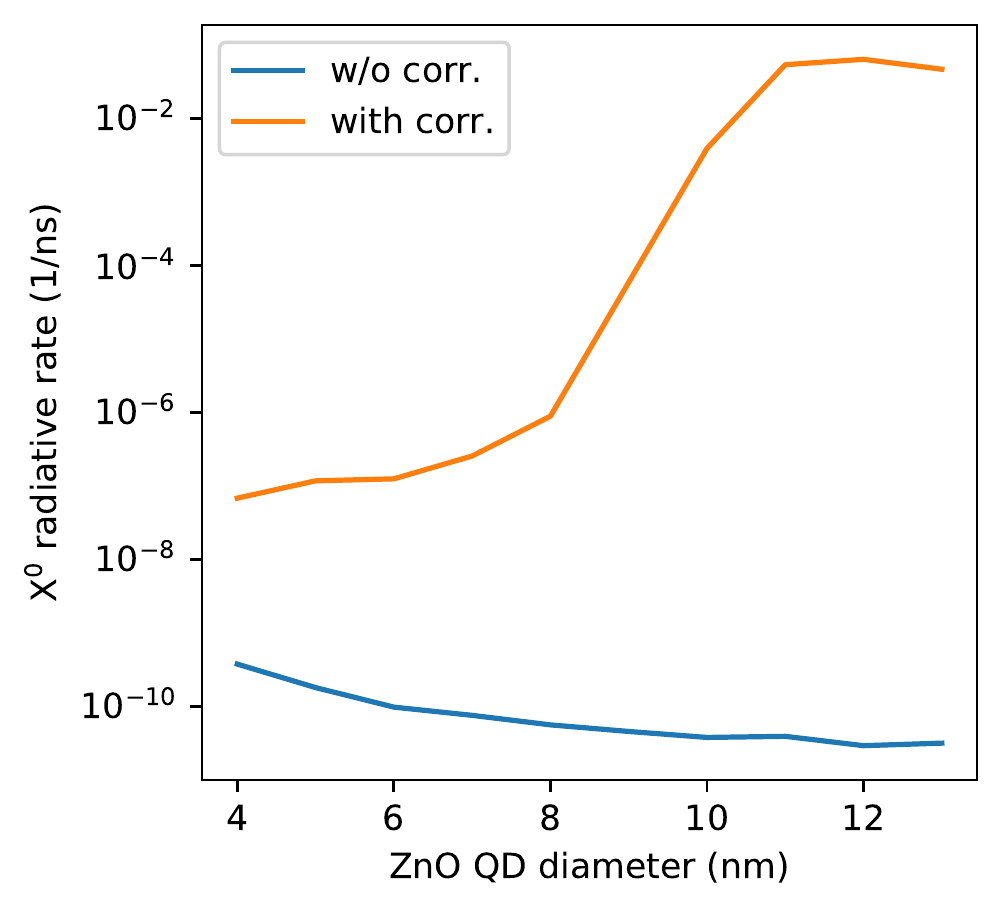}
	\caption{Radiative rate of bright exciton as a function of ZnO QD diameter. The data are shown for calculation using CI with basis of 2~electron and 2~hole SP ground states (w/o corr.) and that for 12 electron and 12 hole SP states (with corr.). Notice the considerable influence Coulomb correlation on the emission rate of ZnO QDs.}
	\label{fig:ZnO_rate}
\end{figure}

Intriguingly, the Coulomb correlation influences sizeably also FSS, BD, and radiative rate of $X^0$~\cite{Kindel2010,Honig2014} as can be~seen in Figs.~\ref{fig:ZnO_FSS}~and~\ref{fig:ZnO_rate}, respectively. Namely, in~Fig.~\ref{fig:ZnO_FSS} the effect of correlation on the exchange interaction even surpasses the multipole expansion of that in ZnO QDs. On the other hand, the overall total magnitude of FSS is found smaller in ZnO QDs compared to that in CdSe/ZnS QDs what is a result of the increased charge separation~\cite{Krapek2015} in the former system.

Similarly, calculating the Fermi's golden rule also for SP excited states higher in energy and including the result of that into the final emission rate, increases the magnitude of the rate by many orders of magnitude compared to that when Fermi's rule is computed only from overlap of the electron and hole SP ground states. Naturally, the emission rate increases with ZnO QD diameter, since in~particular the hole SP states are more closely spaced in~energy with increasing QD size and,~thus, their relative importance in the correlated state of $X^0$ is larger.

\section{Conclusions}

We studied the properties of the electronic structure of~the~exciton in zincblende CdSe/ZnS and wurtzite ZnO quantum dots using a model based on the configuration interaction with basis states obtained from eight-band ${\bf k\cdot p}$ method. The computed emission energies of~the~exciton were verified using the photoluminescence and scanning near-field optical microscopy measurements and with help of a simple model provided correct estimates of CdSe/ZnS dot diameters, which we tested using atomic force microscopy measurements.

The Coulomb correlation was found to have a negligible impact on exciton properties of zincblende CdSe/ZnS dots but turned out to be decisive for both fine structure and emission rate of exciton in wurtzite ZnO quantum dots. 
%
%We identified that the reason for that is the ring-like topology of the ground single-particle hole state, which reduced the overlap with single-particle ground state electrons. 
%
On the other hand, in zincblende CdSe/ZnS dots the dominant mechanism for the exciton formation was the attractive direct Coulomb interaction between electron and hole. Concomitantly, the fine structure in CdSe/ZnS was driven by the multipole terms of the expansion of the exchange interaction, while in ZnO that was again the Coulomb correlation.

\section{Acknowledgments}

Project CUSPIDOR has received funding from the QuantERA ERA-NET Cofund in Quantum Technologies implemented within the European Union's Horizon 2020 Programme. In addition, this project has received national funding from the MEYS and funding from European Union's Horizon 2020 (2014-2020) research and innovation framework programme under grant agreement No 731473. 
The work reported in this paper also was partially funded by projects 20IND05 QADeT, 20FUN05 SEQUME, 17FUN06 Siqust and 20FUN02 POLight that received funding from the EMPIR programme co-financed by the Participating States and from the European Union’s Horizon 2020 research and innovation programme.
%

%\clearpage
%\newpage

\bibliography{references.bib}

\end{document}